\documentclass{rspublic}

\usepackage{epsfig}

\begin{document}

\title{Phase behavior of rod-like virus and polymer mixtures}
\author{Zvonimir Dogic and Seth Fraden}
\affiliation{The Complex Fluid Group, Martin Fisher School of
Physics,
\\Brandeis University, Waltham Massachusetts 02454}
\date{\today}
\maketitle
\begin{abstract}{Depletion Interaction; Isotropic-Smectic Phase
Transition; Hard-rod Fluids}

We have prepared a homologous series of filamentous viruses with
varying contour length using molecular cloning techniques. These
viruses are monodisperse enough to form a stable smectic phase.
Two systems are studied. The first system consists of viruses to
which polymers are covalently attached to the virus surface.
Through studies of the isotropic - cholesteric phase transition we
demonstrate that covalently attached polymers alter the effective
diameter of the virus. Additionally, we have produced mixtures of
viruses whose ratio of effective diameters varies by a factor of
five. The second system is composed of mixtures of rod-like
viruses and non-absorbing polymers. With this system we study the
kinetics of the isotropic - smectic phase transition and describe
observations of a number of novel metastable structures of
unexpected complexity.

\end{abstract}

\vskip2pc
\section{Introduction}
Observation of the nematic phase in aqueous suspensions of
rod-like TMV (tobacco mosaic virus) served as an inspiration for
Onsager to write his seminal paper on the isotropic-nematic (I-N)
phase transition in hard rods (Onsager 1949). Ever since then
biopolymers (DNA, TMV, {\it fd}) have served as an important model
system of hard rods and have often been used to test the Onsager
theory and its various extensions (Meyer 1990; Fraden 1995;
Livolant 1991). In section~\ref{modelsystem} of this paper we
briefly outline the advantages of using the semi-flexible rod-like
{\it fd} or closely related M13 virus as a model system of hard
rods. We demonstrate that using standard procedures of molecular
cloning it is possible to construct genetically modified viruses
with widely varying contour length. These viruses are monodisperse
enough to form a stable smectic phase. In
section~\ref{attachedpolymer} we outline the synthesis of a {\it
fd}-polymer complex and show that polymers covalently attached to
the virus effectively increase the diameter of the rods. By
changing the ionic strength it is possible to observe the
crossover from the regime where the rods are electrostatically
stabilized to where they are stabilized by repulsion between
attached polymers. This synthesis is a convenient way to alter the
diameter of the rod and enables us to study bi-disperse rod
suspensions with different diameters. In section \ref{freepolymer}
we outline the phase behavior of mixtures  {\it fd} virus with
non-absorbing polymer. In particular we focus on the
isotropic-smectic phase transition and describe a number of
different pathways in which the smectic phase nucleates and grows
out of an isotropic suspension of viruses.

\section{{\it fd} virus as a versatile model system of hard rods}
\label{modelsystem}

TMV and {\it fd} viruses form, in order of increasing
concentration of rods, a stable isotropic, nematic or cholesteric,
and smectic phase (Wen \textit{et al.}; Dogic \& Fraden 1997;
Dogic \& Fraden 2000). These two experimental colloidal systems
are the only ones that follow the sequence of liquid crystalline
phase transitions that have been predicted by the theory and
computer simulations of hard rods (Bolhuis \& Frenkel 1997; Vroege
\& Lekkerkerker 1992). Paucity of systems exhibitng smectic phases
is presumably due to polydispersity, which is inherently present
in all other polymeric and colloidal experimental systems due to
the fact that they are chemically synthesized. In contrast to
chemical synthesis, Nature uses DNA technology to produce viruses
that are identical to each other, which results in highly
monodisperse viruses. This high monodispersity of virus
suspensions is the property that makes them an appealing system to
experimentally study the phase behavior of hard rods.

However, there are several important disadvantages that viruses
have compared to synthethic rod-like polymers. Firstly, although
rod-like viruses have very well defined lengths and diameters
studies of how the phase behavior depends on the length to
diameter ratio are non-existent for virus suspensions. Secondly,
the viruses are charged stabilized and therefore their
interactions are not truly hard rod interactions, but in addition
to steric repulsion, have  a long range soft repulsion. It is
important to note that because of the small diameter of the virus
the range of this electrostatic repulsion is always comparable to
the hard core diameter for the range of ionic strengths for which
the stability of the virus against aggregation is not compromised.
Also, because of its protein structure it is impossible to
dissolve the virus in apolar or weakly polar solvents and preserve
the colloidal stability of the virus. It has also been observed
that the virus aggregates in an ionic solution of multivalent
cations. In this section we show that using standard biological
methods it is possible to alter the contour length of the virus
while preserving the monodispresity of the virus. In the
subsequent section we show that by covalently attaching polymer
onto the virus surface we can alter the effective diameter of the
virus, and we have achieved stability of the virus even in the
presence of multivalent cations. It is our hope that the
introduction of these methods will make the viruses a more
appealing model system with which to study phase behavior of rods.

We note that M13 virus with length (L) diameter (D)
(L/D$\approx$130) and construct M13-$T_{n}$3-15 (L/D$\approx$240)
were used in the studies of the concentration dependence of
rotational diffusion almost 20 years ago (Maguire \textit{et al.)
1980}. However, this potentially powerful method was never pursued
in subsequent studies. M13 virus is genetically almost identical
to {\it fd} and has the same contour length with coat proteins
differing by only a single amino acid; negatively charged
aspartate in {\it fd} (asp$_{12}$) corresponding to neutral
asparagine in M13 (asn$_{12}$) ( Bhattacharjee \textit{et al}
1992). This change in a single amino-acid alters the surface
charge by about 30 percent and M13 can easily be distinguished
from {\it fd} by gel electrophoresis. All our clones have their
origin in M13 virus, which also means that they have lower surface
charge then {\it fd} wild type ({\it wt}) system.

Since all available data indicates that the length of the virus is
linearly proportional to the length of the DNA contained in the
virus, the virus length can be extended by simply  introducing
foreign DNA into M13 wt DNA using restriction
endonucleases~(Herrmann \textit{et al.} 1980). However, during
large scale preparation we found that the mutant virus would often
quickly revert to it's wild type form by deleting the foreign DNA.
Another disadvantage of this method is that it is impossible to
construct clones that are shorter then M13 wt. Because of these
reasons we used a well documented phagemid method to prepare our
rod-like viruses with variable contour length (Maniatis \textit{et
al.}  1989). This method allows us to grow clones that are both
longer and shorter then M13 wt. The disadvantage of the phagemid
method is that the helper phage M13KO7 (a virus with contour
length 1.2 $\mu$m) is always present in the final suspension. The
volume fraction of the helper phage depends on the bacterial host
and can vary from 20\% (E. Coli. JM 101) to 5\% (E. Coli. XL-1
Blue). Typically, 0.5 - 1 gram of purified virus can be obtained
in one to two weeks of work. We found that it is possible to
separate the clones from the 1.2 $\mu$m long helper phage by
adjusting the concentration of the bi-disperse, purified virus
suspension such that it is in I-N coexistence. There is a strong
fractionation effect at the I-N transition for bidisperse rods,
with large rods almost entirely dispersed in nematic phase as is
predicted by the theory (Lekkerkerker \textit{et al.} 1984 ,Sato
\& Teramoto 1994 ). Therefore, by keeping only the portion of the
suspension in the isotropic phase we can obtain rods with higher
monodispersity.

All of the viruses grown using the phagemid method are
monodisperse enough to form stable smectic phases as is
illustrated in Fig.~\ref{Fig1}. We note that the measured spacing
of the smectic phase ($\lambda$) is almost identical to the
contour length ($L$) for all the mutants studied. The qualitative
trend that flexibility decreases the smectic layering has been
predicted theoretically and observed experimentally (Dogic \&
Fraden 1997, Polson \& Frenkel 1997,van der Schoot 1996, Tkachenko
1996). Unfortunately, the theories are not accurate enough to be
able to quantitatively predict dependence of smectic spacing on
the flexibility of the rod. We expect that the persistence length
($P \sim 2.2 \mu$m) of all our clones is the same, because all
clones have the same structure and only vary in length. Since the
contour length varies, so to does the ratio of contour to
persistence length $L/P$. Thus we expected that the shorter rods
($L = 0.4 \mu$m) would be relatively stiffer than the longer ones
($L = 1.4 \mu$m) and consequently predicted that the layer spacing
would increase for shorter rods. This was not the case as we
observed that for all lengths the ratio $\lambda / L \sim 1$.

We also find out that {\it fd wt}  (Fig 1c) consistently forms a
smectic phase at a lower concentration then M13 constructs. This
is perhaps explained by the difference in surface charge between
M13 and {\it fd} and the breakdown of the concept of effective
diameter at high concentrations. The {\it fd wt}  is more charged
than M13 and therefore the highly concentrated aligned rods in
nematic phase repel each other more strongly, which results in a
higher effective concentration and thus the nematic-smectic phase
transition occurs at a lower number density of rods. Note that at
low concentrations changing the surface charge by 30\% has
negligible effect on the effective diameter and the phase behavior
of the isotropic - nematic transition (Fig. 1 in Tang and Fraden
(1995)).

With the availability of rods with different contour length we are
able to experimentally explore a number of important issues
pertaining to the phase behavior of hard rods. For pure rods we
can address the question of how flexible can a particle be and
still form a smectic phase. Another important question is the
relative stability of columnar and smectic phase as a function of
rod bidispersity or polydispersity (Bates \& Frenkel 1998, Bohle
\textit{et al.} 1996, Stroobants 1992, van Roij \& Mulder 1996,
Cui \& Chen 1994). For mixtures whose lengths are different enough
there is also a prediction of microseparated smectic phase (Koda
\& Kimura 1994). So far there are no experimental studies on these
subjects, but with our system we can prepare artificially
polydisperse and bidisperse suspensions to explore  these issues.

\section{{\it fd} virus with covalently attached polymer}
\label{attachedpolymer}

Besides preparing viruses with varying contour length we are also
able to alter the effective diameter of the virus by coating it
with the polymer. The amino terminal group of each coat protein of
{\it fd} and M13 virus is exposed to the solution. Through this
chemical site we are able to covalently attach water soluble
polymer Poly(ethylene glycol) (PEG) to the surface of the virus.
End functionalized PEG molecules that readily attaches to amino
groups were obtained from Shearwater polymers. The chemical
reaction was carried out in 100 mM phophate buffer at pH 7.5 for
30 minutes and the virus concentration was kept at 1 mg/ml. For
SSA-PEG-5000 the concentration of PEG was kept the same as the
concentration of the virus in the reaction vessel while for
SPA-PEG-20000 the concentration was four-fold the virus
concentration. The reaction product ({\it fd}-PEG) was separated
from unreacted PEG polymer by repeated centrifugation at 200,000
g. The pellet contained the nematic phase of the {\it fd}-PEG
complex. We diluted a few samples to the concentration of the
isotropic-nematic phase co-existence, and after an exceedingly
long time (up to few months), we observed macroscopic phase
separation. The measured width of the co-existing concentrations
did not differ from the measured widt in {\it fd wt},  which is
about 10\% (Tang \& Fraden 1995). This is an indication that the
absorbed polymer does not significantly alter the flexibility of
the rod-like particles. We infer this from the well established
fact that the width of the I-N coexistence is very sensitive to
the flexibility of the rod (Chen 1993). If we had observed
widening of the I-N coexistence it would have been an indication
that polymer effectively increases rigidity of the rod. Because of
the extremely long time required for complete phase separation, in
order to obtain the points in Figure 2 we diluted the nematic
phase until there was no more birefringence observed. We presume
that this concentration is equal to the concentration of rods in
isotropic phase coexisting with the nematic phase.

To interpret the data in Figure 2 we need to introduce the concept
of the effective diameter ($D_{\mbox{\scriptsize eff}}$). The
isotropic-nematic phase transition for very long rods can be
described at the level of the second virial coefficient, as was
first recognized by Onsager (Onsager 1949). The prediction of the
theory is that the isotropic phase becomes unstable when following
condition is satisfied: $ c\: \pi L^2 D / 4 = 4$, where $c$ is rod
number density, while $L$ and $D$ are the length and the diameter
of the rod. In the same paper Onsager showed how to incorporate
the effect of long range repulsion due to surface charge by
exchanging the bare diameter with an effective diameter
$D_{\mbox{\scriptsize eff}}$, which can be rigorously calculated
and is roughly equal to the distance between two rods where the
intermolecular potential is equal to thermal energy of 1
k$_{\mbox{\scriptsize b}}$T. At high ionic strength
$D_{\mbox{\scriptsize eff}}$ approaches the bare diameter, while
at low ionic strength $D_{\mbox{\scriptsize eff}}$ is much larger
then the bare diameter, and is typically several Debye screening
lengths. The condition for the instability of the isotropic phase
for charged rods becomes $ c \: \pi L^2 D_{\mbox{\scriptsize eff}}
/ 4=4$. It follows that the bare rod number density
 at the I-N phase transition is inversely proportional to
$D_{\mbox{\scriptsize eff}}$. This is experimentally observed for
{\it fd wt} over a wide range of ionic strengths as shown with
square symbols in Fig.~\ref{Fig2}. The full line, which contains
no adjustable parameters, is the numerical solution of the I-N
transition for semiflexible rods where $D_{\mbox{\scriptsize
eff}}$ is calculated by an extension of the Onsager theory (Chen
1993).

Water at room temperature is a good solvent for PEG polymers,
which approximate Gaussian  coils. Thus PEG coated surfaces
interact with each other through long range repulsion (Devanand \&
Selser 1992, Kuhl \textit{et al.) 1994}. Therefore in our {\it
fd}-PEG system, in addition to the already present electrostatic
repulsion between the charged virus surfaces, we introduce
repulsion due to the attached PEG molecules. We expect that for
polymers with large molecular weight and/or at high ionic strength
the dominant interparticle interaction, and consequently
$D_{\mbox{\scriptsize eff}} $ is completely determined by the
polymer diameter because the ionic double layer is confined deep
within the attached polymer. The opposite is true at low ionic
strength and/or low molecular weight polymer. This is exactly the
behavior that is shown in Figure 2. For {\it fd}-PEG-20,000 we
observe that for ionic strengths greater than 2 mM the I - N
transition is independent of ionic strength. This implies that
$D_{\mbox{\scriptsize eff}}$ for the {\it fd}-PEO-20,000 system is
determined entirely by polymer repulsion. The effective diameter
of the particle can be extracted from the I-N co-existence
concentrations since we have shown that there is a relationship
between the effective diameter and concentration of virus:
$c~\mbox{[mg/ml]} = 222/D_{\mbox{\scriptsize eff}}~\mbox{[nm]}$.
For {\it fd}-PEG-5,000  the I - N transition changes from being
dominated by polymer stabilization at high ionic strength to
electrostatic stabilization below 20 mM ionic strength. Because
this transition from polymer dominated to electrostatic dominated
repulsion occurs at a higher ionic strength for {\it fd}-PEG-5,000
compared to {\it fd}-PEO-20,000, the effective diameter of {\it
fd}-PEG-5,000 is smaller than that for {\it fd}-PEO-20,000. The
formula relating the molecular weight ($M_w$) of PEG to it's
radius of gyration ($R_g$) is $R_g=0.215 M_w^{0.583}$~\AA
(Devanand \& Selser 1992). From Figure 2 we can calculate that
{\it fd}-PEG-20000 system has $D_{\mbox{\scriptsize eff}} =
45$~nm, which is approximately approximately equal to
$D_{\mbox{\scriptsize bare}}+4R_g=35$~nm. {\it fd}-PEG-5000
complex has $D_{\mbox{\scriptsize eff}} = 17$~nm at high ionic
strength, while $D_{\mbox{\scriptsize bare}}+4R_g = 19$~nm. This
suggests the model of the polymer being a sphere of radius $R_g$
attached to the surface of the virus, although we expect that the
polymer is deformed by the virus to some extent. In principle, if
the exact shape of the repulsive interaction between two polymer
covered cylindrical surfaces is known, and if the number of
attached polymers per virus is measured, it would be possible to
theoretically calculate the phase diagram for rods with attached
polymers and compare it to experimental findings. However, we have
not yet developed a method to accurately measure the polymer
surface coverage.

We can use our system of rods with different diameters to study
some basic problems in the physics of colloidal liquid crystals.
To prepare a binary mixtures of rods with different diameters we
simply mix {\it fd wt}  and {\it fd}-PEO. The ratio of the
diameters is equal to the ratio of the concentrations at which
these two systems undergo the I-N transition. An additional
advantage of this system is that this ratio can be altered in
continuous way by simply adjusting ionic strength. From
Fig.~\ref{Fig2} it is possible to deduce that at 200 mM ionic
strength the {\it fd}-PEO complex has effective diameter about 5
times thicker then {\it fd} wt. We have observed both
isotropic-isotropic and nematic-nematic demixing in binary
mixtures of {\it fd}-PEG-20000 and {\it fd} wt. Comparison to
available theories is currently underway (Roij \& Mulder 1998,
Sear \& Mulder 1996). In summary, a combination of molecular
engineering and post-expression chemistry has resulted in the
production of gram quantitites of monodisperse rods varying in
length from $0.4 - 1.4 \mu$m and diameter 10 - 50~nm.

\section{Phase behavior of {\it fd} wt virus with non-absorbing polymer}
\label{freepolymer}

Onsager has shown  how to describe the I-N transition of hard rods
with large L/D ratio using the virial expansion of free energy.
The second virial expansion quantitatively describes very long and
thin rods at the isotropic - nematic co-existence, but fails for
highly aligned and concentrated rods. As explained in the previous
section, even systems that have soft repulsion can be successfully
described by the Onsager theory. The reason for this is that the
lowest energy state occurs when two charged rods are perpendicular
to each other. Therefore, charge reduces alignment of the rods,
which in turn increases the accuracy of the virial expansion
(Stroobants \textit{et al.} 1986).  In contrast, if there is
attraction between rods, then perfectly parallel rods are the
configuration with the lowest energy. Consequently, attractions
increase the overall alignment of the rods in a nematic suspension
and decrease the accuracy with which the virial expansion
describes the system. It was shown that for even very slightly
attractive rods the third virial coefficient is almost as large as
the second one (van der Schoot \& Odijk 1992). Currently there is
a lack of both experiments and theories describing the I-N
transition in suspensions of attractive rods and our understanding
of the phase behavior of rods with attraction is rather limited.

We should note that there is a recent theory that introduces
attractions to the study of the I-N transition of hard rods
indirectly by considering mixtures of hard rods and polymers (
Lekkerkerker \& Stroobants 1994). The polymers induce an effective
attraction between colloidal rods through the well known mechanism
of depletion attraction (Asakura \& Oosawa 1958). An advantage of
this system to study the influence of attractions on hard rods is
that it is possible to control both the range of attraction by
varying the molecular weight of  added polymer and the interaction
strength by altering the polymer concentration. However, there is
an important difference between a hard rod/polymer mixture and
suspension of pure rods with attractive interactions because for
the latter the polymer concentration is different across the
coexisting phases and therefore the strengths of attraction
between rods in the isotropic and nematic phases are different (
Lekkerkerker \textit{et al.} 1992).

In our experimental studies of mixtures of {\it fd wt} and
polymers we seek polymers which do not interact with the virus.
The two polymers we use for this purpose are Poly(ethylene glycol)
(PEG) and Dextran. To measure the I-N phase coexistence we mix
concentrated {\it fd} virus and Dextran (M. W. 148,000), dilute
the sample with buffer until two phase co-existence is initiated,
and let the sample phase separate at room temperature, which takes
about two weeks for the slowest phase separating sample. The $R_g$
of 148,000 Dextran is about 11 nm (Nordmeier 1993, Senti
\textit{et al.} 1955). In order to measure the concentration of
both rods and polymers in the coexisting isotropic and nematic
phases we use fluorescently labeled FITC-Dextran. After
appropriate dilution the concentrations of both polymer and {\it
fd} is measured on the spectrophotometer. The resulting phase
diagram is shown in Fig.~\ref{Fig3}. At low polymer volume
fraction the coexisting I-N concentrations  change little from the
pure virus limit and there is little polymer partitioning between
the coexisting phases. At higher polymer volume fractions the
phase diagrams ``opens up" and we measure the coexistence between
a polymer-rich rod-poor isotropic phase and a polymer-poor highly
concentrated nematic phase. The qualitative features in such a
phase diagram are very similar to the theoretically predicted
phase diagram ( Lekkerkerker \& Stroobants 1994, Bolhuis
\textit{et al.} 1997). In a forthcoming publication we will
present detailed experiments of the effects of
 ionic strength, polymer nature, and molecular weight on the phase
diagram (Dogic \textit{et al.} 2001).

When the phase diagram ``opens up'',  the concentration of the
rods in the nematic phase coexisting with the isotropic phase
dramatically increases. For the ionic strength of 100 mM {\it fd}
virus forms a stable smectic phase at 160 mg/ml (Dogic \& Fraden
1997) so it is reasonable to expect a stable isotropic-smectic
(I-S) phase coexistence to supersede the I-N transition for high
enough polymer volume fraction. Indeed, experiments show that
there is a stable I-S coexistence. Since the size of our virus
allows us to visualize individual smectic layers with an optical
microscope we can observe the nucleation and growth of the smectic
phase out of an isotropic suspension in real time. Observation of
typical structures and their temporal evolution are summarized in
the remainder of this paper. All the following images were taken
with a Nikon optical microscope using DIC optics equipped with a
60x water immersion lens and condenser.

The typical I-N tactoid that forms in the two phase region of pure
hard rods, or rods with weak attractions (i. e. low polymer volume
fraction) is shown in Figure~\ref{Fig4}a. The nematic phase
appears as a bright droplet elongated along the nematic director
with a dark background of isotropic rods. In the picture, the rods
are parallel to the plane of the paper and tend to align parallel
to the I-N boundary. As the polymer concentration is further
increased we initially observe nematic droplets as shown in
Fig.~\ref{Fig4}a, but after a few minutes the droplets begin to
change their morphology. Figures 4a to 4k were all taken from the
same sample  and show the time evolution of an initially smooth
tactoid during the first 20-30 minutes of phase separation. In Fig
4b we observe a thin helical sheet wrapped around the nematic
tactoid. The width of the sheet along the direction of the tactoid
is about 1 $\mu$m. We assume that this sheet is a single smectic
layer of rods parallel to the direction of the nematic tactoid
that has nucleated on the nematic surface. This smectic layer
continues to grow and becomes thicker as shown in the side views
of the tactoid in Figs 4c and 4d. Figure 4e shows the same helical
structure, but this time viewed from above (the alignment of the
rods is perpendicular to the paper). We observe that the helical
smectic layer can close itself to form a single ring around the
nematic tactoid. A typical example of this structure is shown in
Fig. 4f where the rods are pointing out of the paper, and in Fig.
4g where rods are parallel to the paper. Two striped tactoids with
smectic rings can coalesce (Fig. 4h) to form droplets with a
variable number of smectic rings as is shown in Fig. 4i, 4j and
4k. Figure 4k to 4m are taken at increasing volume fraction of
polymers. From these three figures we observe that with increasing
polymer concentration the thickness of the smectic rings increases
in comparison to the size of the nematic core. The striped nematic
droplets encircled with smectic layers will proceed to coalesce
until they sediment to the bottom of the sample and reach a size
that is many tens of microns. It should also be noted that not all
tactoids have the closed ring structures, but some instead have a
helical structure that has a beginning and a end. This has
important consequences for the further progress of phase
separation as is demonstrated in Fig.~\ref{Fig5}.

After the sample has been phase separating for few hours we
observe a new kind of structure shown in Fig. 5a. These are
filaments of {\it fd} that have a cross section of $1 \mu$m, which
corresponds to a one particle length. The director is oriented
perpendicular to the fiber axis and precesses in a helical fashion
as in a cholesteric. This results in the helical structures
observed in optical micrographs. The connection between the
twisted sheets and the striped tactoids from Fig.~\ref{Fig4}
coexisting in the same sample is clearly shown in Fig. 5c. The
twisted strands grow slowly out of the smectic rings and over a
period of few days the strands are able to reach lengths of
several hundred microns. We should note that the twisted strand is
a metastable structure with a pronounced tendency to untwist over
a period of days or as one moves along the length of the strand
away from its root at the striped I-N droplet. For example Fig. 5b
to 5g where all taken from the sample sample and show very
different degrees of twisting. Two strands can also connect with
each other as is shown in Fig. 5f. The twisted strands can quite
often form a helical superstructure. Figure 5e is focused onto the
bottom and Fig. 5g is focused on the top of such a structure.
Perhaps such a structure has it's origin in a striped tactoid (Fig
5c) that has for some reason lost its nematic core.

After a few months, as the sample further evolves towards
equilibrium  we observe a number of large sheets that are one rod
length thick. We believe that these are essentially large smectic
membranes. Using the microscope we photograph a sequential series
of images in the plane of focus (xy plane, Fig. 6a) evenly spaced
at 0.2 $\mu$m intervals in the z dimension and from this
information we reconstruct the structure of the membrane in three
dimensions. Fig. 6c shows the image of the membrane perpendicular
to the alignment of the rods from which we deduce that the
diameter of the membrane is about 10 $\mu$m. The cuts through the
xy and yz planes are uniformly one micron thick along the y
direction.

In another series of experiments we studied a mixture of {\it fd}
virus and PEG polymer (M. W. 35,000, $R_g= 9.6$ nm) shown in
Fig.~\ref{Fig7}. The concentration of rods (10 mg/ml) was lower
than in the Dextran/virus mixture described previously, but the
ionic strength was again 110~mM. We increased polymer
concentration until we observed slight turbidity in our sample
indicating the onset of two-phase coexistence. The structures we
observed under these conditions with PEG/virus mixtures are very
similar to the structures observed in Dextran/virus mixtures
illustrated in the previous three figures.  As we increased the
polymer concentration further, we observed a direct formation of
the smectic membrane out of isotropic suspension, instead of their
growth from the striped nematic tactoid. An image of such a
membrane, where all the rods point out of the surface of the paper
is shown in Fig. 7a. The side view (not shown) indicates that the
membrane is essentially one rod length thick. The membranes are
stable over a period of hours, which is surprisingly long. If the
sample is observed for long enough it is possible to observe the
process of coalescence of two smectic membranes. Fig. 7e shows
such a process in a sequence of frames spaced 1/30 seconds apart.
In the first frame the rods in both membranes are aligned in the
same direction. Once the membranes are aligned, the process of
coalescence is complete in about 0.16 seconds.

As the concentration of the polymer is increased further another
pathway to the formation of the smectic phase is observed. We
presume that this process initially begins with the formation of
the smectic membranes just as the one described in the previous
paragraph. However, these membranes never reach the size of the
membranes at lower polymer concentration, which coalesce sideways
as is shown in Fig.~\ref{Fig7}e. Instead, while the membranes are
quite small they stack on top of each other to form long filaments
shown in Fig.~\ref{Fig7}c. Within a few seconds after mixing the
sample these filaments form a percolating network, which is self
supporting and does not sediment over time. As is seen in
Figs.~\ref{Fig7}c the thickness of the filament is not uniform,
but varies from one layer to the other. The irregular thickness of
the filaments does not change even if the sample is left to
equilibrate for few days. From this we can conclude that it takes
rods a very long time to diffuse from one layer to another. We
also observe that as the concentration of the polymer is
increased, the thickness of the filament decreases. The formation
of the filaments can be understood in terms of depletion
attraction. Once a single smectic layer grows to a critical size a
lower energy is achieved by stacking two equal diameter membranes
on top of each other rather then by letting two membranes coalesce
laterally. This is because the strength of the attraction between
two surfaces is proportional to the area of the interacting
surfaces.

It is well known that depletion attraction between a colloid and a
wall is much stronger than the attraction between two colloids
(Dinsmore \textit{et al.} 1997; Sear 1998). Because of this, in
parallel to the bulk phase transitions described previously, there
are competing transitions with the surface of the container. Some
of the structures we observe on the surfaces due to the depletion
attraction are shown in Fig~\ref{Fig7}f and~\ref{Fig7}g.
Fig.~\ref{Fig7}f shows a single smectic layer of rods. By focusing
through the layer in z direction we conclude that this layer is
extremely thin (upper limit of 0.2~$\mu$m). Furthermore, these
layers can stack on top of each other as is shown in
Fig.~\ref{Fig7}g.

Up to now, all the experiments were done with  polymers of roughly
the same radius of gyration (Dextran 150,000 has $R_g=11$ nm, PEG
35,000 has $R_g=9.6$ nm) and at same ionic strength (110~mM). When
we decrease the radius of the polymer (PEG 8,000 has $R_g=4.1$ nm)
we still observe two-dimensional membranes which are composed of
parallel rods. However, as is shown in Fig.~\ref{Fig8}, the
membranes assume a hexagonal shape, which strongly implies that
the rods within the membrane are not a two dimensional fluid, but
a two-dimensional crystal. Figures \ref{Fig8}a to \ref{Fig8}d
where taken at the lowest polymer concentration at which the
crystallization was observed. Under these conditions the induction
time for critical nuclei formation as indicated by the turbidity
of the sample is about 30 minutes. A typical image of a 2D crystal
where the rods within the crystal are pointing out of the plane of
the paper is shown in Fig. \ref{Fig8}a, while the side view where
the alignment of the rods is in the plane of the paper is shown in
Fig \ref{Fig8}b. The thermal fluctuations within the crystal are
easily visible under the microscope and the crystal is easily
deformed as is visible in the side view of the crystal. Often,
instead of observing a flat membrane, we observe a membrane with
screw dislocation located at the nucleation center. The images of
such a membrane from the top view and side view are shown in
Fig.~\ref{Fig8}c and~\ref{Fig8}d. In Fig~\ref{Fig8}4c we can
clearly see that the two layers are on top of each other, but if
we focus through in the z direction we observe that these two
layers belong to the same 2d crystal. This is exactly what we
would expect from a crystal that has a screw dislocation.

If we increase the polymer concentration, the induction time
decreases and an image of these post-critical nuclei is shown in
Fig. \ref{Fig8}e. A typical crystal that usually grows overnight
out of this solution is shown in top view in Fig. \ref{Fig8}f,
while Fig. \ref{Fig8}g shows the side view of such a crystal. A
nucleation center that significantly protrudes out of the 2D
crystalline membrane in clearly visible, and sometimes it is even
possible to observe two 2D crystal membranes connected through the
same nucleation center as shown in Fig 8h. It is important to note
that such a nucleation center is visible in every 2D crystalline
membrane and at all polymer concentrations. Two-dimensional
crystals have been observed in rod-like TMV/BSA mixtures (Adams \&
Fraden 1998a) and these crystals also have a clearly visible
single nucleation site protruding. The fact that the structures
observed in PEG/fd and TMV/BSA system are extremely similar
suggest that the features of 2D crystalline membranes summarized
here are generic to any system of rods with short range
attraction. At even higher polymer concentration, the induction
time is unmeasureably short and typical nuclei that are formed
almost instantaneously are shown in Fig. \ref{Fig8}j. The
resulting crystals display almost no thermal fluctuations, are
much smaller than crystals formed at low polymer concentration,
their number density is much higher, and typically their edges are
much sharper and better defined as is shown in Fig. \ref{Fig8}k
and \ref{Fig8}l.

The influence of both polymer concentration and polymer range has
been extensively studied for three dimensional spherical colloids
(Hagen \& Frenkel 1994, Gast \textit{et al.} 1986, Lekkerkerker
\textit{et al.} 1992). The basic parameter that determines the
behavior of the system is the ratio of the range of attraction
between colloids as compared to the range of the effective hard
core repulsion. On the one hand, if the range of attraction is
very short the vapor-liquid phase transition will be metastable
with regards to the vapor-crystal transition for all conditions.
On the other hand, if the range of attraction is sufficiently long
ranged under certain conditions the vapor-liquid transition will
supersede the vapor-crystal phase transition. Our results on the
formation of two dimensional membranes in the polymer/virus
mixtures agree with this general rule. In the mixture of large
polymer (Dextran M. W. 150,000, $R_g=11$ nm) and {\it fd} virus
the attraction is long ranged and we observe a two dimensional
liquid like membrane. It contrast, in the mixture of small polymer
{PEG 8,000, $R_g=4$ nm) where the attraction is short ranged, we
observe a two dimensional crystalline membrane.

\section{Conclusions}
In the first two sections of this paper have demonstrated the
production of monodisperse rod-like {\it fd} and M13 viruses for
which the contour length and effective diameter was systematically
altered. We plan to use these viruses to study smectic phase
formation as a function of contour length to persistence length,
and the use the polymer-grafted {\it fd} to study smectic phase
formation as a function of the range of interparticle repulsion.
We also plan to study the effects of bidispersity and
polydispersity in both diameter and length on the liquid
crystalline phase transitions. In the latter portion of this
paper, we summarized the behavior of virus/polymer mixtures, which
behave as hard rods with an attractive potential. Although the
interactions between rods in polymer solutions is very simple, we
observe a whole range of novel structures of surprising
complexity. These experiments and previous studies on rod/sphere
mixtures (Adams \textit{et al.} 1998) indicate that there is much
that remains to be understood about the phase behavior of such
mixtures.

\section{Acknowledgment}
This research was supported by the NSF Grant. We thank Kirstin Purdy for
preparation of the cloned virus pGT-N28. Additional information, movies, and
photographs are available online at: www.elsie.brandeis.edu

\smallskip{}

\newpage

\begin{figure}
\centerline{\mbox{\epsfig{file=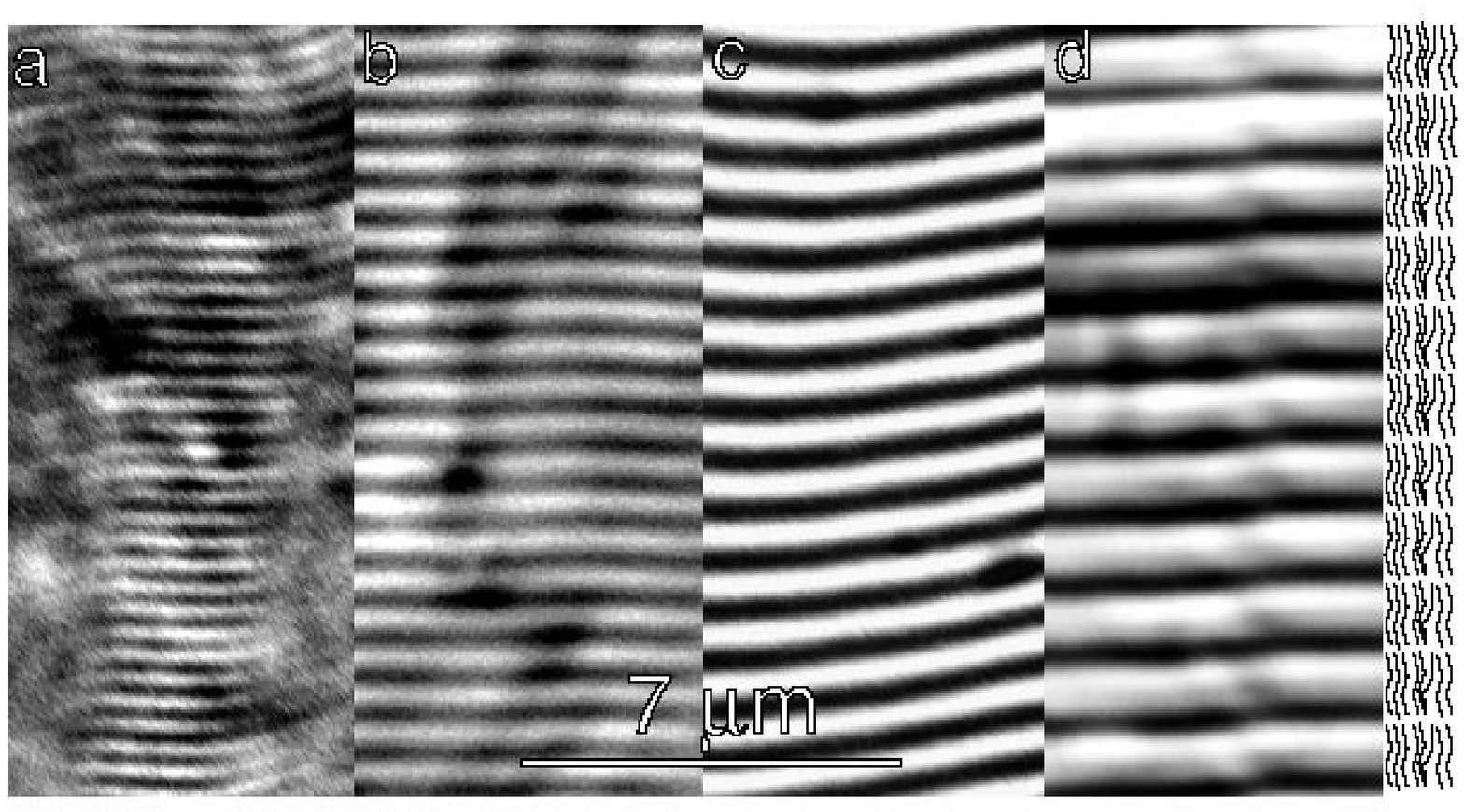,width=5in}}}
\caption{\label{Fig1} Optical micrographs of smectic phases of
three different M13 constructs and {\it fd wt}  (c). The periodic
pattern is due to smectic layers that are composed of two
dimensional liquids of essentially parallel rods, as indicated in
the cartoon on the right. From left to right, the contour length
of the rod-like viruses forming the smectic phase are 0.39 $\mu$m,
0.64 $\mu$m, 0.88 $\mu$m, and 1.2 $\mu$m. The smectic spacing
measured from optical micrographs is 0.40 $\mu$m, 0.64 $\mu$m, 0.9
$\mu$m and 1.22 $\mu$m from image (a) to (d) respectively.}
\end{figure}

\begin{figure}
\centerline{\mbox{\epsfig{file=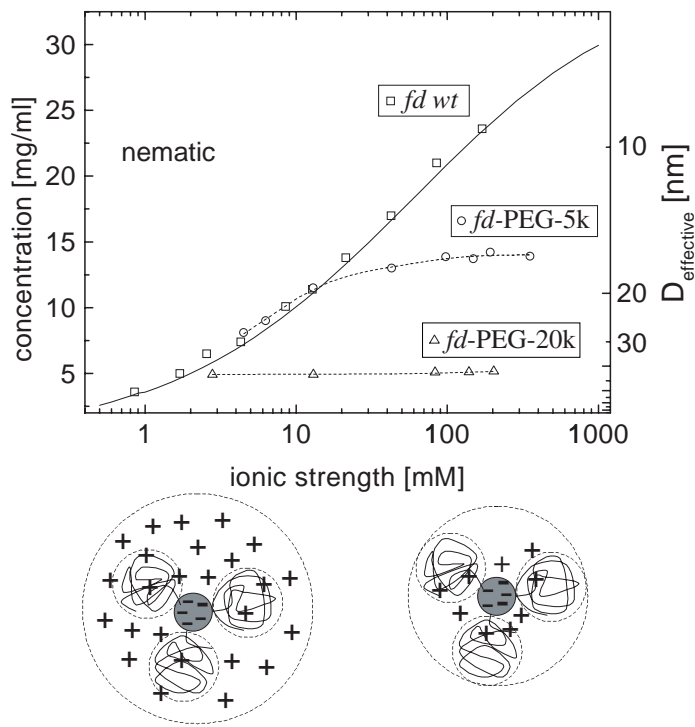,width=4in}}}
\caption{\label{Fig2} Concentration of the virus rods in
coexisting isotropic and nematic phases as a function of ionic
strength and thickness of the PEG layer covalently attached to the
virus. Square points indicate the I-N transition in {\it fd wt}
and were taken from previous work (Tang \& Fraden 1995). The
relationship between the I-N co-existence concentration ($c$) and
electrostatic effective diameter is $c \mbox{[mg/ml]} =
222/D_{\mbox{\scriptsize eff}}$[nm] and is drawn as a solid line.
Circles indicate the I-N transition in {\it fd} coated with
PEG-5,000, while triangles refer to the {\it fd} virus coated with
PEG-20,000. When calculating the concentration of {\it fd}-PEG we
only take into account the {\it fd} core since the polymer density
is not known. The dashed lines are a guide for the eye. At low
ionic strength, electrostatic repulsion determines
$D_{\mbox{\scriptsize eff}}$, while the grafted polymer sets
$D_{\mbox{\scriptsize eff}}$ at high ionic strength, as indicated
in the cartoon of a cross-section of the PEG-virus complex.}
\end{figure}

\begin{figure}
\centerline{\mbox{\epsfig{file=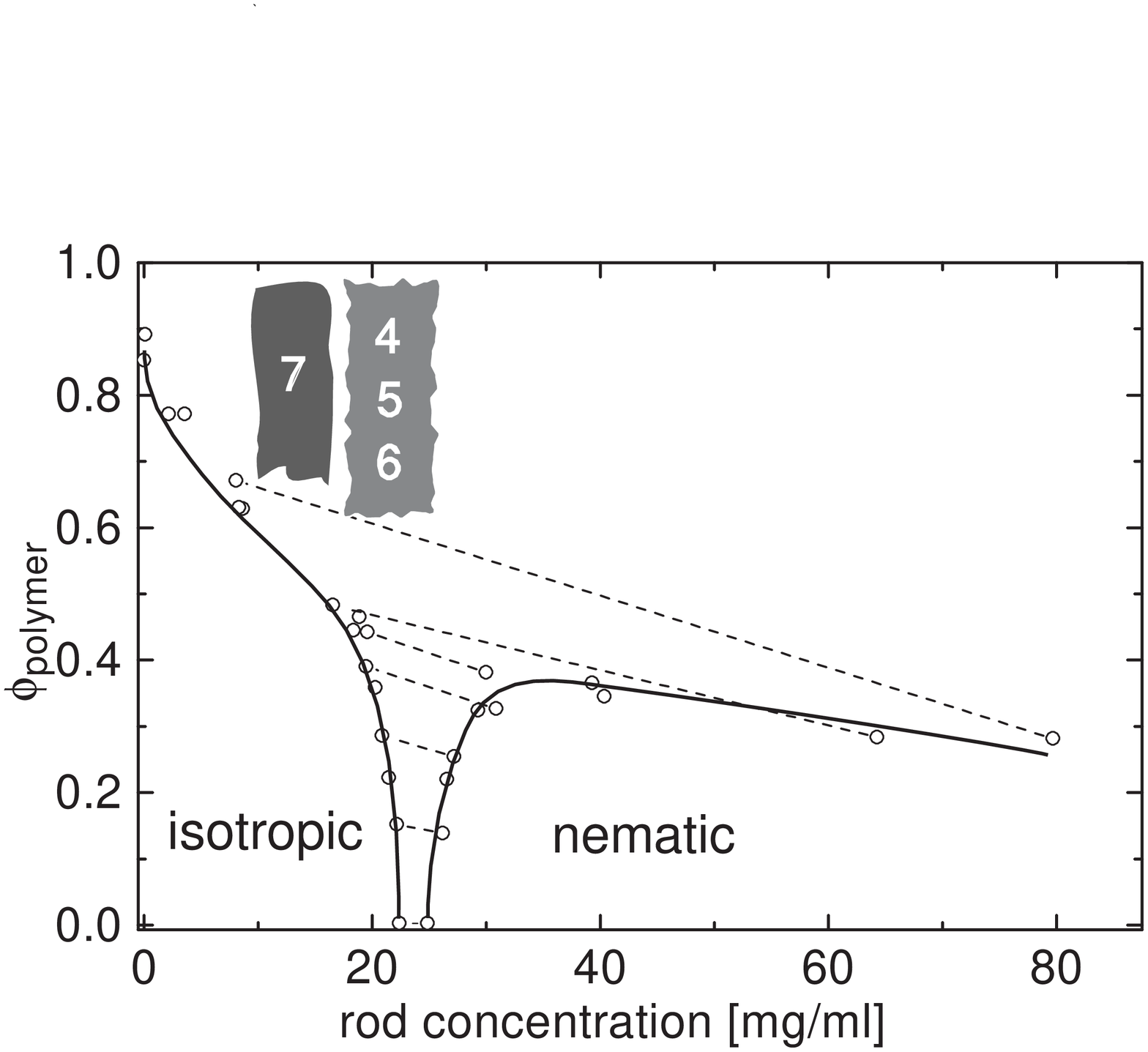,width=4in}}}
\caption{\label{Fig3} Phase diagram of {\it fd wt} and Dextran (M.
W. 148,000) mixture at 200 mM ionic strength. The
R$_{\mbox{\scriptsize g}}$ of Dextran was taken to be 11nm and
vertical axis is given by $\phi_{\mbox{\scriptsize polymers}}=(4
\pi R^3_g/3) (N/V)$ where N/V is the number density of Dextran
polymers. The dashed lines are  tie lines between coexisting
isotropic and nematic phases. Not all of the coexistence lines are
shown for clarity. The full lines are a guide to the eye
indicating the boundary of the two phase region. At high polymer
concentration the rods do not form a uniform phase, but a
percolating network which does not completely sediment and
therefore we are not able to measure its concentration. The region
of the phase diagram labeled ``4'' corresponds to the conditions
of the samples in Figure 4, although the ionic strengths are
different. }
\end{figure}

\begin{figure}
\centerline{\mbox{\epsfig{file=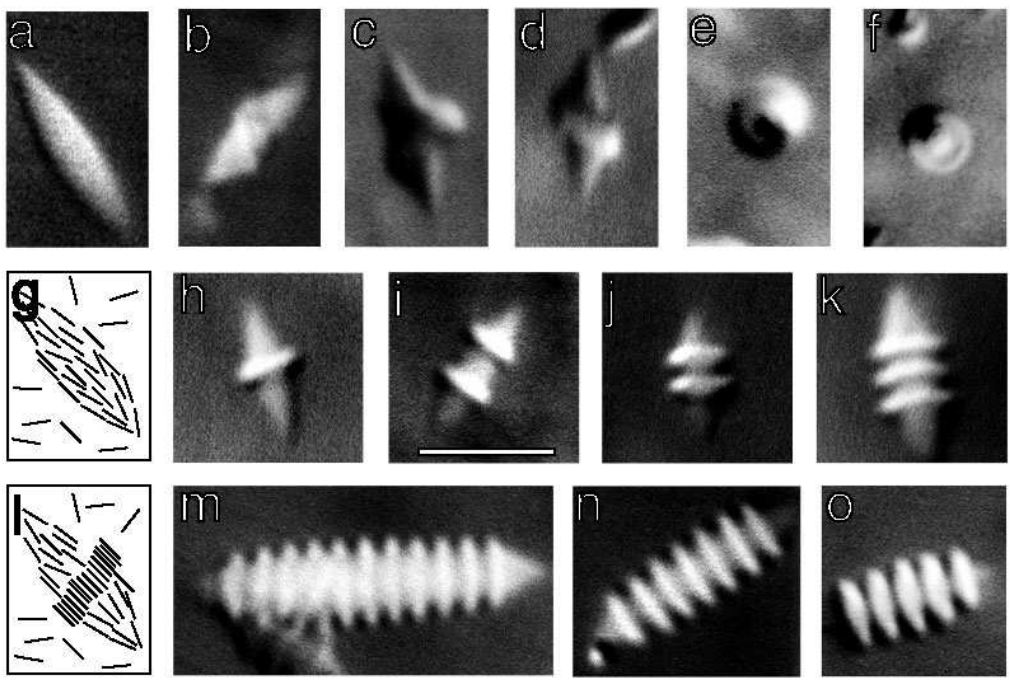,width=5.5in}}}
\caption{\label{Fig4} Initial stages of the phase separation of an
initially isotropic suspension {\it fd} at concentration of 22
mg/ml and Dextran (M. W. 150,000) that shows the formation of
striped tactoids upon addition of Dextran. The ionic strength is
110 mM. The concentration of polymer in picture (a) to (f) and (h)
to (k) is constant and was added to the pure virus suspension
until it became slightly turbid. The concentration of polymer
increases in samples (m) to (o). In figure (g) we sketch the
conformation of rods in typical tactoid at I-N transition in rods
withot attraction. The sketch of the nematic tactoid with the
smectic ring is shown in figure l. The scale bar is 5 $\mu$m long
and all images are taken at same magnification. }
\end{figure}

\begin{figure}
\centerline{\mbox{\epsfig{file=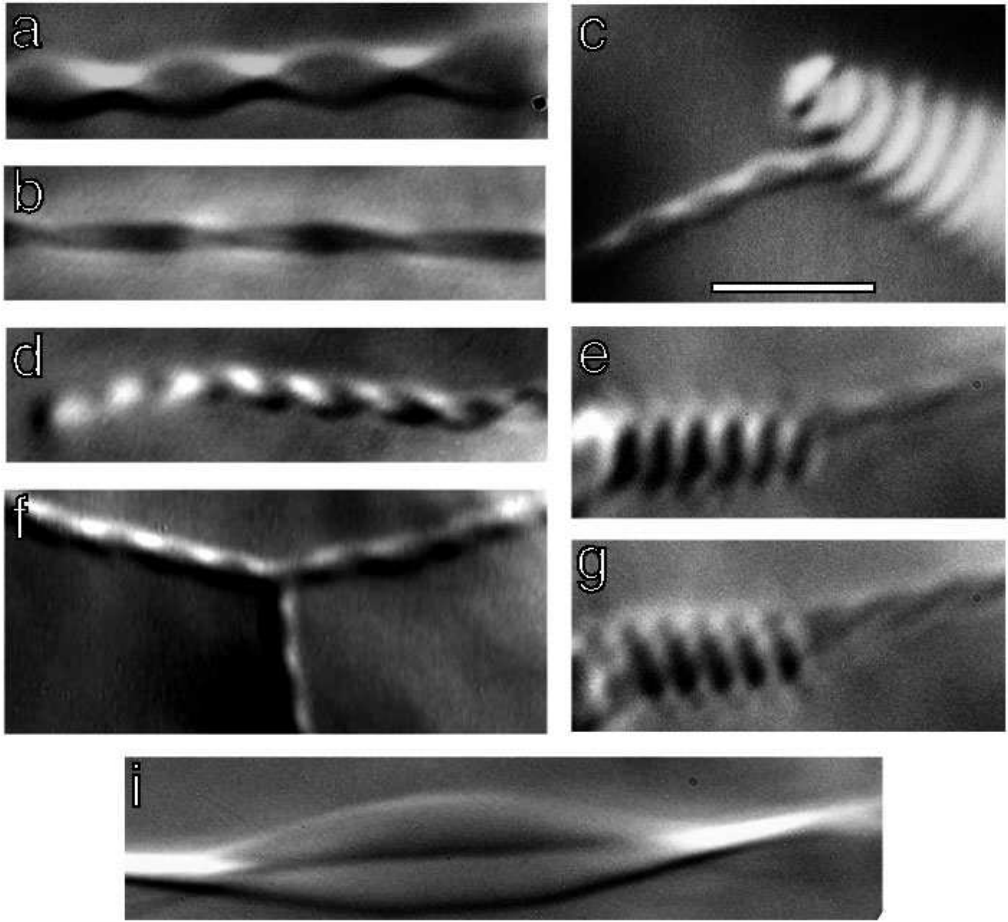,width=5.5in}}}
\caption{\label{Fig5} The twisted strands in (b) to (g) are with
the same conditions as in Figure 4(a). Figure 5(a) is taken at a
higher polymer volume fraction, while figure 5(i) is taken at
lower virus concentration (5 mg/ml). The scale bar indicates 5
$\mu$m.}
\end{figure}

\begin{figure}
\centerline{\mbox{\epsfig{file=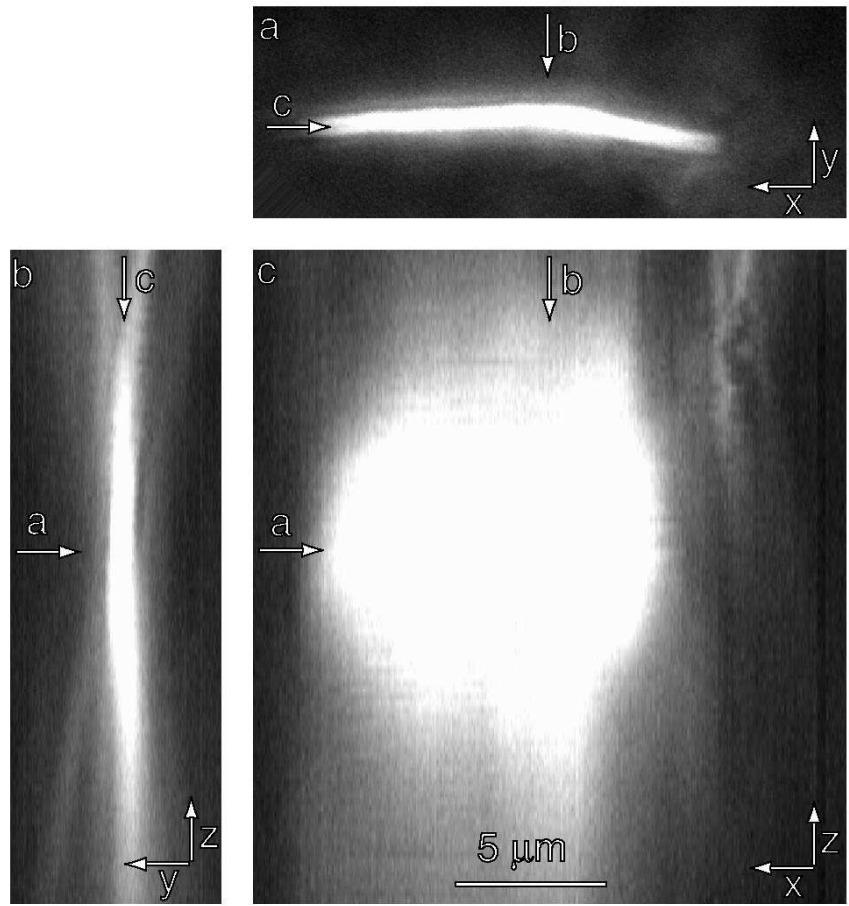,width=5.5in}}}
\caption{\label{Fig6} A three dimensional reconstruction of a
large membrane of a single layer smectic that is observed in a
mixture {\it fd wt} and Dextran 150,000 M.W.  after the it has
been equilibrating for 2 months. Using the microscope a sequential
series of images in the xy plane at different depths z (Fig 6a)
were taken and the image was reconstructed in three dimensions.
Figure 6b shows the image of the membrane cut along the y
direction at the position indicated by  arrow (b) in figure 6(a)
and 6(c).  Equivalently, figure 6c shows the cut of the membrane
perpendicular to the virus axis as indicated by arrow (c) in
figures 6(a) and 6(b). The scale bar indicates 5 $\mu$m.}
\end{figure}

\begin{figure}
\centerline{\mbox{\epsfig{file=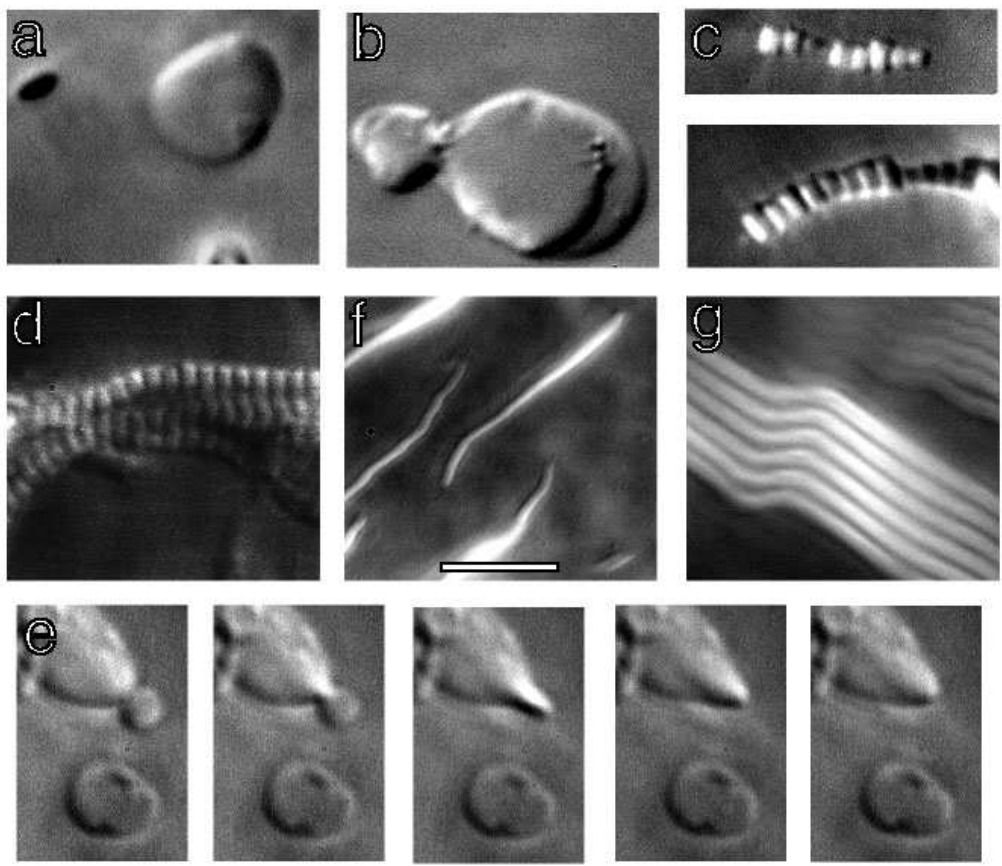,width=5.5in}}}
\caption{\label{Fig7} Phase behavior of mixture {\it fd} and PEG
(M. W. 35,000). At the lowest concentrations of polymer we observe
striped tactoids that are very similar to the ones shown in
previous figures. As the concentration is increased, we observe
formation of a single membrane one rod-length thick that is shown
in figures 7a and 7b. In figure 7e, five successive video frames
spaced 1/30 of seconds apart show coalescence of two smectic
membranes. At an even higher volume fraction of polymer, we
observe filaments shown in figures 7c and 7d that percolate
throughout the entire sample. The phase transitions on the surface
are shown in figures 7f and 7g. The scale bar indicates 5 $\mu$m.}
\end{figure}

\begin{figure}
\centerline{\mbox{\epsfig{file=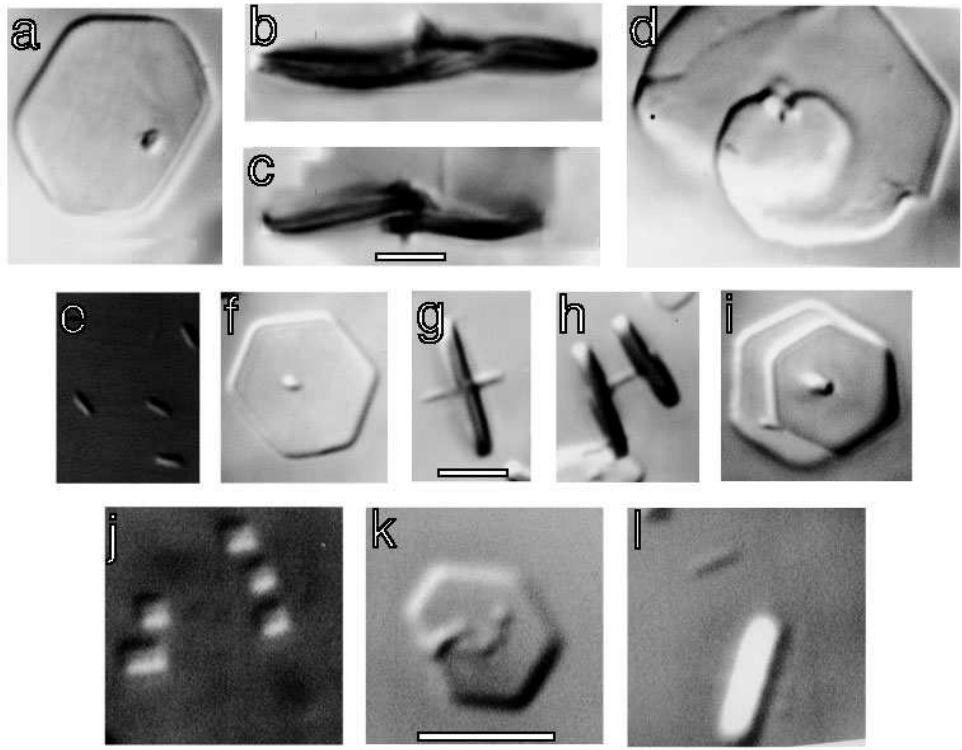,width=5.5in}}}
\caption{\label{Fig8} Optical micrographs of two dimensional virus
crystals observed in mixture of PEG (M.W. 8,000) and {\it fd}
virus at a constant concentration of 15 mg/ml. The first row of
the pictures is at lowest polymer concentration at which the
crystals where observed, the second row is at intermediate polymer
concentration and third row is at highest polymer concentration.
The scale bars are 5 $\mu$m and images in each row are at the same
magnification.}
\end{figure}

\end{document}